\documentclass[prl,
twocolumn,
showpacs,floatfix]{revtex4}
\usepackage{graphicx}
\usepackage{color}
\usepackage{amsfonts}
\usepackage[figuresright]{rotating}  
\usepackage{amssymb}
\usepackage{amsmath}
\usepackage{psfrag}
\usepackage{subfigure}
\usepackage{multirow}
\usepackage{tabularx}
\usepackage{textcomp}
\usepackage{units}

\begin{document}
\def\edd{\epsilon_{\rm dd}}          

\title{Quantum Fluctuations in Dipolar Bose Gases}
\author{Aristeu R. P. Lima}
\affiliation{Institut f\"{u}r Theoretische Physik, Freie Universit\"{a}t Berlin, Arnimallee 14, 14195 Berlin, Germany}
\author{Axel Pelster}
\affiliation{Fachbereich Physik, Universit\"{a}t Duisburg-Essen, Lotharstrasse 1, 47048 Duisburg, Germany}
\date{\today}
\begin{abstract}
We investigate the influence of quantum fluctuations upon dipolar Bose gases by means of the Bogoliubov-de Gennes theory. Thereby, we make use of the local density approximation to evaluate the dipolar exchange interaction between the condensate and the excited particles. This allows to obtain the Bogoliubov spectrum analytically in the limit of large particle numbers. After discussing the condensate depletion and the ground-state energy correction, we derive quantum corrected equations of motion for harmonically trapped dipolar Bose gases by using superfluid hydrodynamics. These equations are subsequently applied to analyze the equilibrium configuration, the low-lying oscillation frequencies, and the time-of-flight dynamics. We find that both atomic magnetic and molecular electric dipolar systems offer promising scenarios for detecting beyond mean-field effects.
\end{abstract}
\maketitle 

\emph{Introduction} -- Bose-Einstein condensates (BEC) with the anisotropic and long-range dipole-dipole interaction (DDI) have received much attention, specially after the condensation of $^{52}$Cr in 2005 \cite{PhysRevLett.94.160401}. This investigation pioneered a series of experiments which led to a robust understanding of the DDI on a mean-field level. Experimental successes include the direct observation of the DDI in the time-of-flight (TOF) dynamics \cite{stuhler_j_2005,strong-pfau}, the stabilization of a purely dipolar gas \cite{stabilization-pfau}, and the observation of a d-wave Bose-nova explosion \cite{d-wave-pfau}. In the meantime, the DDI has also been observed even in $^{87}$Rb \cite{rubidium} and evidences of it have been found in $^{7}$Li \cite{lithium}. Recently, an important experiment has been realized, in which the influence of the DDI upon the oscillation frequencies of $^{52}$Cr has been studied \cite{PhysRevLett.105.040404}. Parallel to these experiments, much theoretical work has been pursued \cite{theory}. For instance, building on the previous construction of the dipolar pseudo-potential \cite{marinescu_m_1998}, a solution of the mean-field Gross-Pitaevskii (GP) equation was obtained \cite{odell_dhj_2004,eberlein_c_2005}, which accounts quantitatively for the static as well as the dynamic properties of trapped dilute $^{52}$Cr gases \cite{stuhler_j_2005,strong-pfau,PhysRevLett.105.040404}. Nowadays, dipolar interactions in magnetic systems are, therefore, considered to be relatively well understood in terms of the GP mean-field theory. Nonetheless, highly magnetic atoms, such as dysprosium \cite{dysp_experiment_two}, or strongly polar heteronuclear molecules, exemplified by $^{40}$K$^{87}$Rb \cite{K.-K.Ni10102008,citeulike:6565167,carr}, are expected to push the understanding of dipolar systems beyond the edge of mean-field theory's domain of validity.

Already in the case of pure contact interaction, ultracold quantum gases have been the scenario for investigating many-body theories. Theoretically, much work has been carried out based on the Lee-Huang-Yang (LHY) correction for the equation of state \cite{lee_td_1957}, when implemented in terms of superfluid hydrodynamics \cite{PhysRevLett.81.4541,braaten}, or on quantum Monte Carlo simulations \cite{PhysRevLett.95.030404}. Experimentally, quantum fluctuations have been studied in setups, which favor their enhancement, such as in optical lattices \cite{PhysRevLett.96.180405}, or in the presence of Feshbach resonances in both bosonic \cite{PhysRevLett.101.135301} and fermionic \cite{PhysRevLett.98.040401} systems. In particular, the importance of the latter experiment should be stressed, as it achieved a precision level which is capable of distinguishing the predictions of Bardeen-Cooper-Schrieffer (BCS) mean-field theory from QMC results for the oscillation frequencies of low-lying modes at the BEC side of the BEC-BCS crossover. This demonstrates unequivocally that collective modes provide an adequate testing ground for many-body theories of interacting quantum gases. In this letter, we theoretically study quantum fluctuations in dipolar Bose gases in the large particle number regime. Within the realm of the Bogoliubov-de Gennes (BdG) theory, we first calculate the condensate depletion and the ground-state energy. Then, by applying superfluid hydrodynamics, we discuss the quantum corrections to both static and dynamic properties of the system focusing on the observability of these effects.

\emph{BdG Theory} -- Consider a Bose gas of $N$ dipoles with mass $M$ in a cylinder symmetric trapping potential $U_{\rm tr}({\bf x})=M\omega_{x}^{2}(x^{2}+y^{2}+{\lambda}^{2}z^{2})/2$ with the trap aspect ratio $\lambda$. The corresponding Hamilton operator reads
\begin{equation}
\hat{H} \!=\! \int\!\!{\rm d}{^{3}x} \hat{\Psi}^{\dagger}({\bf x})\!\left[h_{0} \!+\! \int\!\frac{{\rm d}{^{3}y}}{2}  \hat{\Psi}^{\dagger} ({\bf y})  V_{\rm int}\!\left({\bf x} \!-\!{\bf y}\right)\hat{\Psi}^{}({\bf y})\right]\!\hat{\Psi}^{}({\bf x})
\label{2nd_quant_hamil}
\end{equation}
with $h_{0} = -{\hbar^{2}\nabla^{2}}/{2M} + U_{\rm tr}({\bf x})$. Moreover, $\hat{\Psi}^{\dagger}({\bf x})$ and $\hat{\Psi}^{}({\bf x})$ denote the usual creation and annihilation field operators, respectively. Considering the dipoles to be aligned along the $z$ axis, the interaction potential reads
\begin{equation}
V_{\rm int}({\bf x}) = \frac{4\pi\hbar^{2}a_{s}}{M}\!\left[\delta({\bf x}) \!+\! \frac{3a_{\rm dd}}{4\pi a_{s}|{\mathbf x}|^{3}}\left(1\!-\!3\frac{z^{2}}{|{\mathbf x}|^{2}}\right)\right],
\label{inter_pot}
\end{equation}
where $a_{s}=M g /4\pi \hbar^{2}$ is the s-wave scattering length and $a_{\rm dd}$ denotes a length scale which characterizes the dipolar interaction, irrespective whether it is of magnetic or electric nature. At this point, it is useful to introduce the dimensionless relative interaction strength $\edd=a_{\rm dd}/a_{s}$.

In order to study dipolar Bose gases beyond the mean-field approximation, we follow the Bogoliubov prescription and decompose the field operator according to $\hat{\Psi}^{}({\bf x}) = \Psi_{}({\bf x}) + \delta\hat{\psi}^{}({\bf x})$, where $\Psi_{}({\bf x})$ is the condensate wave function and $\delta\hat{\psi}^{}({\bf x})$ accounts for the quantum fluctuations. By considering only the leading term in this expansion, one obtains the GP theory of dipolar BECs, which was exactly solved in the Thomas-Fermi (TF) approximation \cite{odell_dhj_2004,eberlein_c_2005}. In order to go beyond the GP theory, we insert the Bogoliubov prescription into the grand-canonical Hamiltonian $\hat{K}=\hat{H}-\mu \hat{N}$, with the chemical potential $\mu$ and the number operator $\hat{N}$, and retain terms up to second order in the fluctuations. Then we decompose the fluctuating field operator according to $\delta\hat{\psi}^{}({\bf x}) = {\sum_{\nu}}'\left[{\mathcal U}_{\nu}({\bf x})\hat{\alpha}_{\nu}+ {\mathcal V}_{\nu}^{*}({\bf x})\hat{\alpha}_{\nu}^{\dagger}\right]$. Here, ${\mathcal U}_{\nu}^{}({\bf x})$ and ${\mathcal V}_{\nu}^{}({\bf x})$ are the Bogoliubov amplitudes, whereas $\hat{\alpha}_{\nu}^{\dagger}$ and $\hat{\alpha}_{\nu}^{}$ denote bosonic creation and annihilation operators. The index $\nu$ is associated with excited energy levels and, therefore, the sum excludes the $\nu=0$ mode, which represents the condensate. By requiring the Hamiltonian to be diagonal, we derive the BdG equations
\begin{eqnarray}
\!\!\!\!\int\!\!{\rm d}{^{3}y}\!\!\left(\begin{array}{cc}
\!\!H_{{\mathcal U},{\mathcal U}}\!\left({{\bf x} ,{\bf y}}\right) & \!\!H_{{\mathcal U},{\mathcal V}}\!\left({{\bf x} ,{\bf y}}\right) \\       
\!\!H^{*}_{{\mathcal U},{\mathcal V}}\!\left({{\bf x} ,{\bf y}}\right) & \!\!H^{*}_{{\mathcal U},{\mathcal U}}\!\left({{\bf x} ,{\bf y}}\right)\!\!    \end{array}
\!\!\right)\!\!\left(\begin{array}{c}
\!\!{\mathcal U}_{\nu}^{}({\bf y})\! \\ \!\!{\mathcal V}_{\nu}^{}({\bf y})\!                                             
                                             \end{array}\!\!\right) \!=\! \varepsilon_{\nu}\!\!\left(\begin{array}{c}
\!\!\!{\mathcal U}_{\nu}^{}({\bf x}) \\ \!\!\!-\!{\mathcal V}_{\nu}^{}({\bf x})                                             
                                             \end{array}
\!\!\!\right)\!\!,
\label{BdG_eqs_gen_bec}
\end{eqnarray}
where the diagonal matrix elements are given according to $H_{{\mathcal U},{\mathcal U}}\!\left({{\bf x} ,{\bf y}}\right) \!=\! \delta\!\left({\bf x}\! -\!{\bf y}\right)\!H_{\rm Fl}({\bf y})\!+\Psi^{*}({\bf y})\!V_{\rm int}\!\left({\bf x}\! -\!{\bf y}\right)\Psi^{}({\bf x})$ and the off-diagonal ones according to $H_{{\mathcal U},{\mathcal V}}\left({{\bf x} ,{\bf y}}\right)=\Psi({\bf y})V_{\rm int}\left({\bf x} -{\bf y}\right)\Psi^{}({\bf x})$. In these expressions, an important role is played by the functionally diagonal part, which is given by the fluctuation Hamiltonian
\begin{equation}
H_{\rm Fl}({\bf x}) = h_{0}- \mu + \int{\rm d}{^{3}x'} \Psi^{*}({\bf x}') V_{\rm int}\left({\bf x} -{\bf x}'\right)\Psi^{}({\bf x}').
\label{fluct_hamil}
\end{equation}
Moreover, in the framework of the BdG theory, the oper\-a\-tor $\hat{\alpha}_{\nu}^{}$ annihilates the ground state $|0\rangle$, so that excitations are interpreted as quasi-particles with energy $\varepsilon_{\nu}$.

In general, solving the BdG equations exactly is very difficult and numerically arduous in the case of long-range interactions \cite{PhysRevA.74.013623}. Nonetheless, they provide an adequate starting point for investigating quantum fluctuations in most cases of interest. Typically, the number of particles is large enough, so that we can neglect the condensate kinetic energy and its wave function becomes $\Psi^{}({\bf x})=\sqrt{n_{0}({\bf x})}$, where $n_{0}({\bf x})$ denotes the TF condensate density. Moreover, in this regime, a semiclassical approximation for the excitations is appropriate. Let $q_{\nu}\left({\bf x}\right)$ be either ${\mathcal U}_{\nu}\left({\bf x}\right)$ or ${\mathcal V}_{\nu}\left({\bf x}\right)$. The approximation is implemented through $\varepsilon_{\nu} \rightarrow \varepsilon \left({\bf x},{\bf k}\right)$, $\sum_{\nu} \rightarrow \int{\rm d}^{3}k/(2\pi)^{3}$, and ${q}_{\nu} \rightarrow {q}\left({\bf x},{\bf k}\right)e^{i{\bf k}\cdot{\bf x}}$, where ${q}\left({\bf x},{\bf k}\right)$ is a {\sl slowly} varying function of ${\bf x}$ \cite{giorgini_thermo}. With this, the free Hamiltonian becomes $h_{0}\left({\bf x},{\bf k}\right) = \hbar^{2}{\bf k}^{2}/2M+U_{\rm tr}({\bf x})$. Despite the simplification, we still cannot solve Eqs.~(\ref{BdG_eqs_gen_bec}) because of their non-local character. The semiclassical approximation also offers a procedure to deal with this problem. Consider the exchange term in Eq.~(\ref{BdG_eqs_gen_bec}), which, in general, reads $I_{\rm Ex}  \equiv  \int{{\rm d}^{3}y} H_{{\mathcal U},{\mathcal V}}\left({{\bf x} ,{\bf y}}\right){ q}_{\nu}\left({\bf y}\right)$. Within the {\sl local density approximation} (LDA), it becomes $I_{\rm Ex} \approx q({\bf x},{\bf k}) n_{0}({\bf x}){V}_{\rm int}({\bf k})$,  with ${V}_{\rm int}({\bf k})=g[1+\edd (3\cos^{2}\theta -1)]$ and $\cos\theta=\hat{\bf k}\cdot\hat{\bf e}_{z}$. This is the leading order in a systematic gradient expansion \cite{PhysRevA.55.3645}. To estimate the error of ignoring the next order, we replace the gradients as ${\nabla}_{\bf x}\rightarrow 1/R_{\rm TF},{\nabla}_{\bf k}\rightarrow 1/K_{c}$, with $R_{\rm TF}$ the mean Thomas-Fermi radius and $K_{c}$ the momentum scale defined by the speed of sound. In terms of the condensate density at the trap center, one finds the LDA to be valid for $3\times[N^{2}a_{s}^{3}n_{0}({\bf 0})]^{\frac{1}{6}}\times[1+\edd \left(3\cos^{2}\theta -1\right)]^{\frac{1}{2}}\gg1$. For $\edd<1$, this is commonly met in BEC experiments.

One of the important physical features brought about by quantum fluctuations is the {\sl condensate depletion}, i.e., particles are expelled from the condensate to excited states by interactions. The depletion density is, in general, given by $ n({\bf x})-n_{0}({\bf x}) = \sum_{\nu}'{\mathcal V}_{\nu}^{*}({\bf x}){\mathcal V}_{\nu}^{}({\bf x})$ and reduces in the present semiclassical theory to
\begin{equation}
\Delta n({\bf x}) = \frac{8n({\bf x})}{3}{\mathcal Q}_{3}(\edd )\sqrt{\frac{n({\bf x})^{}a_{s}^{3}}{\pi}}.
\label{deplet_dens_harm}
\end{equation}
Here, $n({\bf x})$ denotes the total density of the gas and the auxiliary function ${\mathcal Q}_{l}(x) = \int_{0}^{1}{{\rm d}u}\left(1-x + 3\,x\, u^{2}\right)^{l/2}$ describes the contribution of the DDI.  
In particular, the function ${\mathcal Q}_{3}$ is an increasing monotonic function of the relative interaction strength $\edd$, which is bounded between ${\mathcal Q}_{3}(0)=1$ and ${\mathcal Q}_{3}(1)=3\sqrt{3}/4\approx1.30$. For $^{52}$Cr, the most extensively studied dipolar system, the magnetic moment of $6\,\mu_{\rm B}$ with the Bohr magneton $\mu_{\rm B}$ and the s-wave scattering length $a_{s}\approx100~a_{0}$ with the Bohr radius $a_{0}$ yield the relative interaction strength $\edd^{}\approx0.16$. Then, the gas parameter $n({\bf x})^{}a_{s}^{3}$ evaluated at the trap center is typically of the order $n({\bf 0})^{}a_{s}^{3}\approx 10^{-4}$ \cite{strong-pfau}. For a homogeneous system with this value of the gas parameter, the fractional depletion amounts to $\Delta N/N\approx1.5\,\%$ and is not significantly affected by the DDI.

The presence of quantum fluctuations also gives rise to a {\sl shift of the ground-state energy}. The present BdG theory yields the following correction to the energy density
\begin{equation}
\frac{\Delta E({\bf x})}{n({\bf x})} = \frac{64}{15} g n({\bf x})^{}{\mathcal Q}_{5}(\edd )\sqrt{\frac{n({\bf x})a_{s}^{3}}{\pi}},
\label{gs_ener_dens_harm}
\end{equation}
where ${\mathcal Q}_{5}$ is also a monotonic function and grows from ${\mathcal Q}_{5}(0)=1$ up to ${\mathcal Q}_{5}(1)=3\sqrt{3}/2\approx2.60$. Thus, physical quantities directly related to the ground-state energy offer much better perspectives for observing dipolar quantum fluctuations. Indeed, with the help of Eq.~(\ref{gs_ener_dens_harm}) a corresponding LHY equation of state for a homogeneous system can be derived, which leads to the Beliaev sound velocity \cite{beliaev2} both in the absence \cite{PhysRevLett.81.4541} and in the presence of the DDI. Note that both functions ${\mathcal Q}_{3}$ and ${\mathcal Q}_{5}$ become imaginary for $\edd>1$. This arises from the fact that a homogeneous dipolar gas is only stable for $\edd\leq1$, otherwise the DDI dominates over the contact interaction and its attractive part leads to collapse.

\emph{ Superfluid Hydrodynamics} -- At $T=0$, the BdG theory states irrespective of the two-particle interaction that the whole sample is superfluid though not all particles are condensed. Thus, the system can be studied by means of the superfluid hydrodynamics. Indeed, by writing the superfluid velocity as ${\bf v}=\nabla\chi$, the superfluid hydrodynamic equations can be obtained by extremizing the action \cite{griffin}
\begin{equation}
{\mathcal A}[n_{},\chi] = -\int{\rm d}{t}{{\rm d}^{3}x}{n_{}}\left\{M\left[\dot{\chi}+\frac{1}{2}\nabla\chi^{2}\right]+e_{\rm}\!\left[n\right]\right\}
\label{action_fact}
\end{equation}
with respect to the density $n({\bf x},t)$ and the velocity potential $\chi({\bf x},t)$. The energy density of the system as a functional of the particle density $e_{\rm}\!\left[n\right]$ contains a mean-field component, which reads in TF approximation
\begin{equation}
e_{\rm MF} = U_{\rm tr}({\bf x})\!+\!\frac{g}{2}n_{}({\bf x},t)\!+\!\int\!\!{{\rm d}^{3}x'}\frac{V_{\rm dd}({\bf x}\!-\!{\bf x'})}{2}n_{}({\bf x'}\!,t),
\label{hamil_fac}
\end{equation}
and a quantum correction, which is given in the case of a dipolar Bose gas by the rhs of Eq.~(\ref{gs_ener_dens_harm}). Let us, now, extremize the resulting action (\ref{action_fact}) within a variational approach. To this end, we adopt a harmonic ansatz for the velocity potential $\chi({\bf x},t) = [\alpha_{x}(t)x^{2}+\alpha_{y}(t)y^{2}+\alpha_{z}(t)z^{2}]/2$ and a parabolic ansatz for the particle density
\begin{equation}
n_{}({{\bf x}},t) = \frac{15N}{8\pi \overline{R}^{3}(t)} \left[1-\sum\limits_{i=x,y,z}\frac{x^{2}_{i}}{R_{i}^{2}(t)}\right]
\label{ansa_dens_var}
\end{equation}
for $n_{}({{\bf x}},t)\ge0$ and $n_{}({{\bf x}},t) =0$ elsewhere. The bar denotes geometric average. Note that the particular density profile in (\ref{ansa_dens_var}) corresponds to the exact mean-field TF solution \cite{odell_dhj_2004,eberlein_c_2005}. This is consistent with the BdG theory, which assumes that interactions are not too strong.

In order to derive quantum corrected equations of motion for the TF radii ${R}_{i}$, we eliminate, at first, the variational parameters for the velocity potential through their equations of motion $\dot{\alpha}_{i}=\dot{R}_{i}/R_{i}$. Then, in the case of a cylinder symmetric trap, we obtain
\begin{eqnarray}
\!\!\!\!\!\!\!\!\!\ddot{R}_{x} & \!\!=\! &\! -\omega_{x}^{2}R_{x} \!+\! \frac{15gN/4\pi}{ M R_{x}\overline{R}^{3}}\!\!\left[1-\edd  A\!\left(\frac{R_{x}}{R_{z}}\right) \!+\! \frac{\beta}{\overline{R}^{\frac{3}{2}}}\right]\!,\nonumber\\
\!\!\!\!\!\!\!\!\!\ddot{R}_{z} & \!\!=\! &\! -\omega_{z}^{2}R_{z} \!+\! \frac{15gN/4\pi}{ M R_{z}\overline{R}^{3}}\!\!\left[1+2\edd  B\!\left(\frac{R_{x}}{R_{z}}\right) \!+\! \frac{\beta}{\overline{R}^{\frac{3}{2}}}\right]\!,
\label{eq_motion_cyl_QF}
\end{eqnarray}
together with the auxiliary functions
\begin{eqnarray}
\!\!\!\!\!\!\!\!\!A\left(x\right) & = & 1 + \frac{3}{2}\frac{{x}^{2}f_{s}\left(x\right)}{{x}^{2}-1},\quad
B\left(x\right) = 1 + \frac{3}{2}\frac{f_{s}\left(x\right)}{{x}^{2}-1}.
\label{aux_func_QF}
\end{eqnarray}
Here, $f_{s}(x)$ denotes the anisotropy function
\begin{eqnarray}
f_{s}(x) & \equiv & \frac{1+2x^2}{1-x^2} - \frac{3 x^2\tanh^{-1}\sqrt{1-x^{2}}}{(1-x^{2})^{3/2}},
\end{eqnarray}
which arises for both BECs \cite{odell_dhj_2004,glaum} and non\-su\-per\-flu\-id fermionic dipolar gases \cite{ourpaper}. In Eqs.~(\ref{eq_motion_cyl_QF}), the quantum correction is governed by the pa\-ram\-e\-ter $\beta = \gamma {\mathcal Q}_{5}(\edd )\sqrt{a_{s}^{3}N}$, with the constant $\gamma = {{3^{\frac{3}{2}}\cdot5^{\frac{3}{2}}\cdot7^{}}/{2^{\frac{13}{2}}}}\approx 4.49$. Consequently, setting $\beta=0$ in Eqs.~(\ref{eq_motion_cyl_QF}) leads to the exact mean-field TF results \cite{odell_dhj_2004,eberlein_c_2005}.

The beyond mean-field equations of motion for dipolar Bose gases (\ref{eq_motion_cyl_QF}) are the main result of the present letter. They allow for investigating both the static and the dynamic properties of these systems. To this end, we solve them by considering $\beta$ as a small perturbation, as this is consistent with the assumption of small fluctuations.

\begin{figure}[t]
\centerline{\includegraphics[scale=.625]{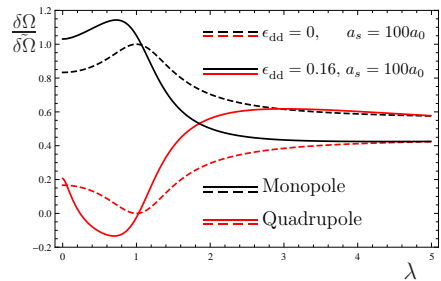}}
\caption{(color online.) Quantum correction to the oscillation frequencies.}
\label{deltaomega_percent}
\end{figure}

\emph{Results} -- As is well known, the DDI affects the system in an anisotropic manner. Thus, in contrast to the case of a Bose gas with contact interaction alone, quantum fluctuations lead to a correction in the gas aspect ratio in equilibrium. To investigate this effect quantitatively, we set the rhs of Eqs.~(\ref{eq_motion_cyl_QF}) to zero and solve them up to first order in $\beta$. Thereby, one finds that the TF radii are given according to $R_{i}=R_{i}^{0}+\delta R_{i}$, where $R_{i}^{0}$ denotes the mean-field value and $\delta R_{i}$ represents the quantum correction. Thus, the beyond mean-field aspect ratio is $\kappa\equiv{(R_{x}^{0}+\delta R_{x}^{})}/({R_{z}^{0}+ \delta R_{z}}) \approx \kappa^{0}\left(1 + \delta\kappa\right)$. Usually, the correction to the aspect ration in the present experiments is quite small. For $^{52}$Cr \cite{strong-pfau}, for example, one has at most ${\delta\kappa}\lesssim10^{-2}$, which is too small to be detected. For stronger interactions, this turns out to be an important effect, since it implies a characteristic dependence on the trap geometry due to the DDI.
Consider, e.g., the experimental values for particle number $N=3\times10^{4}$ and trap frequencies $\overline{\omega}\approx2\pi\times660$~Hz as in Ref.~\cite{strong-pfau}, but instead of chromium values, use $a_{s}=190~a_{0}$ and $\edd=0.7$. This could represent a dysprosium sample, which is heavier than chromium and possesses a larger magnetic moment of $10~\mu_{\rm B}$ yielding the dipolar length $a_{\rm dd}\approx 133~a_{0}$. Then, the correction for the aspect ratio would be about ${\delta\kappa}\approx27~\%$ for moderate cigar-like traps.

Let us, now, analyze the low-lying excitations. To this end, we apply for the TF radii the ansatz $R_{i}(t) = R_{i}(0) + \eta_{i}\sin\left({\Omega t}+\varphi\right)$ and obtain by linearizing Eqs.~(\ref{eq_motion_cyl_QF}) a matrix equation which leads to the frequencies of the monopole and of the quadrupole modes up to first order in $\beta$. In Fig.~\ref{deltaomega_percent}, we plot the resulting corrections to the oscillation frequencies as functions of the trap aspect ratio $\lambda$ in units of $\tilde{\delta\Omega} = ({63\sqrt{\pi}}/{128})\sqrt{a_{s}^{3}~n({\bf 0})}$ for $\edd=0.16$. The dashed lines represent the Pitaevskii-Stringari result for contact interaction alone \cite{PhysRevLett.81.4541}, which we immediately recover by setting $\edd=0$. Fig.~\ref{deltaomega_percent} shows a dependence on the trap aspect ratio, which is substantially different from the contact case and stems from the quantum correction for the aspect ratio. Thus, tuning the trap aspect ratio could be used to identify the effects of the DDI at the many-body level in ultracold Bose gases. This, however, requires stronger interactions. Indeed,  the oscillation frequency of the intermediate mode of $^{52}$Cr has recently been investigated in a triaxial trap \cite{PhysRevLett.105.040404}. For that experiment we estimate the quantum correction of the oscillation frequencies to be of the order $\tilde{\delta\Omega}\approx 3\times10^{-3}$, due to the small particle number and weak trap. For more favorable values of these parameters, such as those in Ref. \cite{strong-pfau}, one already obtains $\tilde{\delta\Omega}\approx 1~\%$. For the previous hypothetical dysprosium setup, one could even have $\tilde{\delta\Omega}\approx 4~\%$, which should render these effects observable.

\begin{figure}[t]
\centerline{\includegraphics[scale=.75]{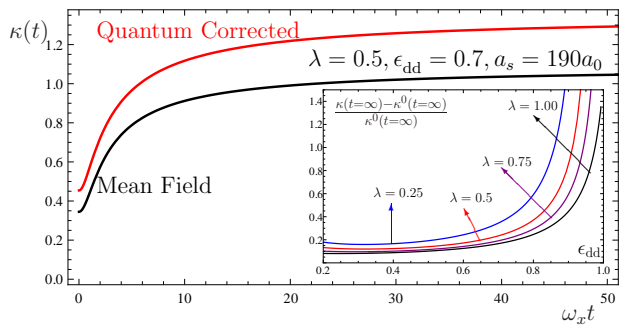}}
\caption{(color online.) Aspect ratio as a function of time.}
\label{deltaTOF}
\end{figure}

For the TOF dynamics, the quantum correction of the equilibrium TF radii also plays an important role. Indeed, by inserting the expansion $R_{i}(t)=R_{i}^{0}(t)+\delta R_{i}(t)$, one derives a set of coupled differential equations for the mean-field TF radii $R_{i}^{0}(t)$ and their corrections $\delta R_{i}(t)$. For a system consisting of $^{52}$Cr atoms, no signal of quantum fluctuations can be seen in the present TOF analysis and the mean-field theory provides again a good agreement with the experimental results \cite{stuhler_j_2005,strong-pfau}. For stronger interactions, however, the situation changes. In Fig.~\ref{deltaTOF}, we plot the ratio $\kappa(t)={[R_{x}^{0}(t)+\delta R_{x}^{}(t)]}/{[R_{z}^{0}(t)+ \delta R_{z}(t)]}$ as a function of time for the previous dysprosium setup. Then, clear differences with respect to the mean-field TOF dynamics show up, specially at large times. To investigate the dependence of the asymptotic aspect ratio on the system parameters we have settled the dipolar length at $a_{\rm dd}\approx 133~a_{0}$ and have varied the s-wave scattering length $a_{s}$, so as to vary the relative strength from $\edd=0.2$ up to $\edd=1.0$. Our results are shown in the inset of Fig.~\ref{deltaTOF}, where the asymptotic value $[\kappa(\infty)-\kappa^{0}(\infty)]/\kappa^{0}(\infty)$ is plotted against $\edd$ for different trap aspect ratios. The results reasonably deviate from the mean-field values, so that quantum fluctuations should be observable in the TOF dynamics of Bose gases of highly magnetic atoms.

\emph{Conclusion} -- Applying the BdG theory in the limit of large particle numbers, we obtained analytic expressions for the condensate depletion and the ground-state energy of dipolar Bose gases. This allowed us to study key properties of the system such as low-lying excitations and TOF dynamics beyond the mean-field approximation. We find that both polar molecules and highly magnetic atoms are realistic scenarios for observing unprecedented beyond mean-field dipolar physics.

We thank J. Dietel and L. Santos for useful discussions and the DFG for financial support (KL256/53-1).

\bibliographystyle{unsrt}

\end{document}